\documentclass[two column,prl,showpacs]{revtex4}
\usepackage{amsfonts}
\usepackage{amsmath}
\usepackage{amssymb}
\usepackage{graphicx}
\providecommand{\U}[1]{\protect\rule{.1in}{.1in}}

\begin{document}
\title{Magnetic anisotropy and reversal in epitaxial Fe/MgO(001) films revisited}
\author{Qing-feng Zhan,$^{1}$ Stijn Vandezande,$^{1}$ Kristiaan Temst,$^{2}$ and Chris
Van Haesendonck$^{1}$}
\affiliation{$^{1}$Laboratorium voor Vaste-Stoffysica en Magnetisme and Institute for
Nanoscale Physics and Chemistry (INPAC), Katholieke Universiteit Leuven,
Celestijnenlaan 200\thinspace D, BE-3001 Leuven, Belgium}
\affiliation{$^{2}$Instituut voor Kern- en Stralingsfysica and Institute for Nanoscale
Physics and Chemistry (INPAC), Katholieke Universiteit Leuven, Celestijnenlaan
200 \thinspace D, BE-3001 Leuven, Belgium}


\begin{abstract}
We investigate the magnetization reversal in Fe/MgO(001) films with fourfold in-plane magnetic anisotropy and an additional uniaxial anisotropy
whose orientation and strength are tuned using different growth geometries and post growth treatments. The previously adopted mechanism of
180$^{\text{o}}$ domain wall nucleation clearly fails to explain the observed 180$^{\text{o}}$ magnetization reversal. A new reversal mechanism
with two successive domain wall nucleations consistently predicts the switching fields for all field orientations. Our results are relevant for
a correct interpretation of magnetization reversal in many other epitaxial metallic and semiconducting thin films.

\end{abstract}
\maketitle

Magnetic anisotropy is one of the most important properties of metallic and semiconducting thin-film magnets~\cite{RPG-59-1409, AM-19-323}. In
magnetic films of cubic systems an in-plane fourfold magnetic anisotropy is expected, but often an additional uniaxial magnetic anisotropy (UMA)
is observed to be superimposed on top of the fourfold anisotropy~\cite{PRL-93-117203, PRL-94-137210}. The extra UMA has been attributed to
different origins, including a self-shadowing effect occurring during oblique deposition~\cite{APL-91-092502, EPL-75-119, PRL-98-046103}, the
bonding between film and substrate~\cite{PRB-63-193301, PRL-95-137202}, and the N\'{e}el surface effect on a stepped
substrate~\cite{PRL-68-1212, PRB-65-184419}. Moreover, ion sputtering has been demonstrated as a reliable tool to control the orientation and
strength of UMA~\cite{PRL-91-167207, PRL-96-057204}.

When applying thin films for magnetic data storage and spintronic devices, the magnetization reversal mechanisms and their dependence on the
anisotropy symmetry need to be known and controlled. The magnetization reversal process for combined cubic and uniaxial anisotropies is
sensitive to the specific anisotropy geometry and strength~\cite{PRB-51-15964, APL-90-062109}. Depending on the field orientation, hysteresis
curves with one and two steps are observed in various films, and explained in terms of nucleation and propagation of $90^{\text{o}}$ and
$180^{\text{o}}$ domain walls (DWs)~\cite{JAP-78-7210, PRB-72-054407}. A model based on minimizing the magnetic energies has been introduced
with DW nucleation energies $\epsilon_{90^{\text{o}}}$ for $90^{\text{o}}$ DWs and $\epsilon _{180^{\text{o}}}$ for $180^{\text{o}}$ DWs,
respectively, in order to account for the observed switching fields~\cite{JAP-78-7210}. A special magnetic switching process involving three
steps can be observed when the additional UMA along the cubic easy axis exceeds $\epsilon_{90^{\text{o}}}$~\cite{PRL-79-4018, PRB-63-104431}.
Until now, such a switching has been assumed to be mediated by two $90^{\text{o}}$ DW nucleations at the first and the third step and one
$180^{\text{o}}$ DW nucleation occurring in between.

Previous work on Fe/MgO(001) films grown at normal incidence revealed a weak UMA along Fe$\left\langle 010\right\rangle $ ~\cite{JMMM-210-341}.
Recently, we successfully relied on ion sputtering to manipulate the strength of the in-plane UMA along Fe$\left\langle 110\right\rangle $ in
Fe/MgO(001)~\cite{APL-91-122510}. Park \textit{et al.} found that a pronounced UMA can be induced in Fe/MgO(001) by relying on oblique-incidence
molecular beam epitaxy (MBE) growth~\cite{APL-66-2140}. Up to now, all of the measured hysteresis loops in Fe/MgO(001) only revealed one or two
steps.

Here, we report on a detailed study of magnetization reversal in Fe/MgO(001) films, where strength and orientation of UMA are tuned either by
ion sputtering or by oblique-incidence MBE growth. A novel mechanism is introduced with two successive DW nucleations to explain the
$180^{\text{o}}$ magnetic switching that occurs for one-step and three-step loops. Our model consistently explains the experimental results for
films with different UMA, revealing the universal nature of the magnetization reversal.

In general, the in-plane UMA, which is superimposed on the cubic anisotropy $K_{1}$ of Fe, can be separated into two components: $K_{u1}$ along
[010] and $K_{u2}$ along [110]~\cite{NJP-9-354}. If $K_{u1}\ll K_{1}$ and $K_{u2}<K_{1}$, the component $K_{u2}$ rotates the position of the
overall easy axes backwards with respect to the uniaxial hard axis over an angle $\delta$ that is approximately given by
$\delta=\frac{1}{2}\sin^{-1}(K_{u2}/K_{1})$~\cite{PRB-51-15964}, as illustrated in Fig.~1(a). In order to obtain a UMA with different
orientation and strength, three Fe films were grown by MBE on MgO(001) substrates using different growth conditions and post growth treatments.
For sample A, the incident Fe beam was at an angle of $49^{\text{o}}$ with respect to the surface normal and with azimuthal angle along Fe[010].
During deposition of sample B at normal incidence, the substrate was rotated around the surface normal. The nominal thickness of samples A and B
was $15\,\mathrm{nm}$, as monitored by a calibrated quartz crystal oscillator. The growth geometry for sample C was the same as for sample B,
but the nominal thickness was $100\,\mathrm{nm}$. Subsequently, sample C was sputtered with $2\ \mathrm{keV}$ Ar$^{+}$ ions at an incidence
angle of $60^{\text{o}}$ with respect to the surface normal and with azimuth fixed in between Fe[100] and Fe[1$\overset{-}{1}$0]. After
sputtering for 250 minutes, the film thickness was reduced to $15\,\mathrm{nm}$, as verified by \textit{ex situ} x-ray reflectometry. Before
removing the samples from the vacuum chamber, they were capped with a $2\,\mathrm{nm}$ thick protective Au layer. The magnetic properties were
measured by the longitudinal and transverse magneto-optical Kerr effect (MOKE) for different field orientation $\phi$ as defined in Fig.~1(a).

For sample A, a considerable UMA along [010] is introduced by the oblique deposition. Three-step loops as well as one-step and two-step loops
are observed at different $\phi$, as illustrated in Figs.~1(b) to 1(d). The switching events, which occur for increasing field and $0^{\text{o}}
<\phi<90^{\text{o}}$, are [$\overset{-}{1}$00]$\rightarrow$[100] for the one-step loops, [0$\overset{-}{1}$0]$\rightarrow$[$\overset{-}{1}
$00]$\rightarrow$[010] for the two-step loops, and [0$\overset{-}{1} $0]$\rightarrow$[$\overset{-}{1}$00]$\rightarrow$[100]$\rightarrow$[010]
for the three-step loops~\cite{PRL-79-4018}. The magnetization switches by $180^{\text{o}}$ for the one-step loops and for the second step of
the three-step loops, and by $90^{\text{o}}$ for the other steps. The corresponding spin orientations are marked by the arrows that are enclosed
in a square in Figs.~1(b) to 1(d).

Up to now $90^{\text{o}}$ as well as $180^{\text{o}}$ DW nucleations have been invoked to interpret the $90^{\text{o}}$ and $180^{\text{o}}$
magnetic transitions, respectively~\cite{JAP-78-7210, PRL-79-4018}. The coercivity related to the DW nucleation energy can be derived from the
energy gain between the local minima at the initial and final easy axes involved in the transition~\cite{JAP-78-7210}. The theoretical switching
fields are obtained as $H_{c1}=(\epsilon_{90^{\text{o}}}-K_{u1})/[M(\sin\phi-\cos\phi)]$ for the magnetic switching process
[0$\overset{-}{1}$0]$\rightarrow$[$\overset{-}{1}$00], $H_{c2}=(\epsilon_{90^{\text{o}}}+K_{u1})/[M(\sin\phi+\cos\phi)]$ for
[$\overset{-}{1}$00]$\rightarrow$[010], $H_{c3}=(\epsilon_{90^{\text{o}}}+K_{u1})/[M(\sin\phi-\cos\phi)]$ for [100]$\rightarrow$[010], $H_{c4}
=(\epsilon_{90^{\text{o}}}-K_{u1})/[M(\cos\phi-\sin\phi)]$ for [010]$\rightarrow$[100], and $H_{c}=\epsilon_{180^{\text{o}}}/[\left\vert
2M(\cos\phi)\right\vert ]$ for [$\overset{-}{1}$00]$\rightarrow$[100], where $M$ is the magnetization.

For sample A, the $\phi$ dependence of the measured switching fields can be nicely fitted using the theoretical switching fields (see
Fig.~2(a)), provided we assume the switching fields correspond to $H_{c1}$, $H_{c2}$ and $H_{c3}$ (see Figs.~1(c) and 1(d)). The fitting results
in the parameters $K_{u1}/M=2.70\pm0.02\,\mathrm{mT}$ and $\epsilon_{90^{\text{o}}}/M=0.61\pm0.02\,\mathrm{mT}$, where
$K_{u1}>\epsilon_{90^{\text{o}}}$ is the necessary condition for the occurrence of three-step loops~\cite{PRL-79-4018}. Following
Refs.~\cite{JAP-78-7210, PRL-79-4018} we try to describe the experimental switching field $H_{c180^{\text{o}}}$, which corresponds to a
$180^{\text{o}}$ magnetic transition, in terms of $180^{\text{o}}$ DW nucleation. Using the corresponding theoretical switching field $H_{c}$,
we obtain the fit that is shown in the two insets to Fig.~2(a). $H_{c}$ reaches a minimum at $\phi=0^{\text{o}}$ and a maximum at
$\phi=90^{\text{o}}$. The experimental switching field $H_{c180^{\text{o}}}$ reveals, however, a peak at both $\phi=0^{\text{o}}$ and
$90^{\text{o}}$. Moreover, around $\phi =90^{\text{o}}$ $H_{c}$ has a slope that disagrees with experiment. $180^{\text{o}}$ DW nucleation
clearly fails to describe the $180^{\text{o}}$ magnetic transition in sample A. Surprisingly, the theoretical expression for $H_{c2}$, which
corresponds to a $90^{\text{o}}$ DW nucleation, allows us to fit the $H_{c180^{\text{o}}}$ data (see below for more details).

The universal character of the absence of $180^{\text{o}}$ DW nucleation for Fe/MgO(001) is confirmed by the $\phi$ dependence of the switching
fields for samples B and C in Figs.~2(b) and 2(c). Only one-step and two-step loops are observed. As discussed in our recent
publication~\cite{APL-91-122510}, the switching route for the two-step loops appearing in samples B and C
($180^{\text{o}}-\delta\rightarrow90^{\text{o}}+\delta\rightarrow-\delta$ for $-\delta<\phi<45^{\text{o}}$) is different from the path for
sample A. When $\delta=0$, i.e., $K_{u2}=0$, the experimental switching fields for this type of two-step loop should correspond to the above
derived expressions for $H_{c2}$ and $H_{c4}$. In case $\delta\neq0$, i.e., $K_{u2}\neq0$, the theoretical expressions need to be extended:
\[
H_{c2}=\frac{\epsilon_{90^{\text{o}}-2\delta}+K_{u1}(\cos^{2}\delta-\sin
^{2}\delta)}{M[\cos(\phi+\delta)+\sin(\phi-\delta)]}\;,
\]
\[
H_{c4}=\frac{\epsilon_{90^{\text{o}}+2\delta}-K_{u1}(\cos^{2}\delta-\sin
^{2}\delta)}{M[\cos(\phi+\delta)-\sin(\phi-\delta)]}\;,
\]
where $\epsilon_{90^{\text{o}}-2\delta}$\ and $\epsilon_{90^{\text{o}}+2\delta}$ are the corresponding DW nucleation energies.

For sample B the two switching fields ($H_{c2}$, $H_{c4}$) have a dependence on $\phi$ that is symmetric about $\left\langle 100\right\rangle $.
Moreover, the angular dependence of $H_{c2}$ reveals a clear, abrupt step when crossing $\left\langle 110\right\rangle $, as illustrated in
Fig.~2(b). We conclude that sample B has a small in-plane UMA along [010]~\cite{JAP-78-7210}. The fitting parameters are
$K_{u1}/M=0.19\pm0.01\,\mathrm{mT}$ and $\epsilon _{90^{\text{o}}}/M=0.36\pm0.01\,\mathrm{mT}$. Because $K_{u1}<\epsilon _{90^{\text{o}}}$,
three-step loops cannot be observed.

In sample C, UMA with components along both [010] and [110] is introduced by the Ar$^{+}$ ion sputtering. The overall easy axes are observed to
deviate from $\left\langle 100\right\rangle $ by an angle $\delta=3^{\text{o}}$, i.e., $K_{u2}/K_{1}\approx0.1$. From the results in Fig.~2(c)
we find that the UMA component along [010] and the DW nucleation energies are $K_{u1}/M=1.69\pm0.02\,\mathrm{mT}$,
$\epsilon_{90^{\text{o}}-2\delta}/M=1.83\pm 0.02\,\mathrm{mT}$, and $\epsilon_{90^{\text{o}}+2\delta}/M=2.29\pm 0.02\,\mathrm{mT}$,
respectively. Because $K_{u1}$ is comparable to $\epsilon_{90^{\text{o}}-2\delta}$, the path $270^{\text{o}}+\delta
\rightarrow-\delta\rightarrow90^{\text{o}}+\delta$ is energetically more favorable when compared to the counterclockwise path via $180^{\text{o}
}-\delta$ for the whole range of angles $45^{\text{o}}<\phi<135^{\text{o}}$. Consequently, both $H_{c2}$ and $H_{c4}$ change monotonously within
this range.

We again try to describe $H_{c180^{\text{o}}}$ for the one-step loops of samples B and C in terms of $180^{\text{o}}$ DW nucleation, as
illustrated in the insets to Figs.~2(b) and 2(c), respectively. The theoretical curves clearly disagree with experiment. From the fitting for
our samples with different anisotropy geometry and strength we conclude that $90^{\text{o}}\pm2\delta$ magnetic transitions in Fe/MgO(001) films
are mediated by $90^{\text{o}}\pm2\delta$ DW nucleation, but $180^{\text{o}}$ magnetization reorientations are not mediated by $180^{\text{o}}$
DW nucleation.

Which mechanism dominates the $180^{\text{o}}$ magnetic reversal? Based on the obtained values for $K_{u1}$ and $\epsilon_{90^{\text{o}}}$, we
plot the energy difference between the relevant easy axes as a function of the applied field in Fig.~3(a) for sample B at $\phi=10^{\text{o}}$
and in Fig.~3(b) for sample A at $\phi=65^{\text{o}}$, respectively. In the previously adopted model, the energy difference between
$180^{\text{o}}$ and $0^{\text{o}}$ is treated in terms of \textit{one single barrier}, and $180^{\text{o}}$ DW nucleation occurs at $H_{c}$
when $\Delta E_{180^{\text{o}}\rightarrow 0^{\text{o}}}=\epsilon_{180^{\text{o}}}$, where $\epsilon_{180^{\text{o}}}$ is assumed to correspond
to $2\epsilon_{90^{\text{o}}}$~\cite{PRL-79-4018}. According to our analysis the switching between [$\overset{-}{1}$00] and [100] is governed by
\textit{two separate energy barriers} between [$\overset{-}{1}$00] and [010] and between [010] and [100], respectively. The switching then
corresponds to two $90^{\text{o}}$ DW nucleation processes. The energy barrier for the transition from [$\overset{-}{1}$00] and [010] becomes
$\Delta E_{180^{\text{o}}\rightarrow90^{\text{o}}}=\epsilon_{90^{\text{o}}}$ at $H_{c2}$. However, since $\Delta
E_{90^{\text{o}}\rightarrow0^{\text{o}}}$\ already exceeds $\epsilon_{90^{\text{o}}}$ at $H_{c2}$, the domains along [010] are unstable and
cannot grow. Therefore, a second nucleation of domains along the final [100] remanent direction occurs at $H_{c2}$, and the two successive
$90^{\text{o}}$ DW nucleations appear as one single step in the MOKE\ loops. In case $\delta\neq0$, this process consists of a
$90^{\text{o}}-2\delta$ DW nucleation and a subsequent $90^{\text{o}}+2\delta$ DW nucleation, or vice versa. Based on our new model the
experimental switching fields $H_{c180^{\text{o}}}$ for all three samples are fitted by the expressions for $H_{c2}$ in Figs.~2(a) to 2(c) and
the insets. This way, all switching fields can be nicely fitted by consistently using the same $\epsilon_{90^{\text{o}}\pm2\delta}$ and $K_{u1}$
values for the complete range of angles.

In case of two successive DW nucleations, $H_{c4}$ is not an experimental observable switching field. $H_{c4}$ only indicates $\Delta
E_{90^{\text{o}}+\delta\rightarrow-\delta}=\epsilon_{90^{\text{o}}+2\delta}$. We have plotted in Figs.~2(a) and 2(b) the \textquotedblleft
virtual\textquotedblright \ $H_{c4}$ values for samples A and B. When $0^{\text{o}}<\phi<45^{\text{o}}$, the two successive DW nucleations
appear as one-step loops for $H_{c2}>H_{c4}$ (see Fig.~3(a)). For $H_{c2}<H_{c4}$, the magnetization loops reveal a two-step behavior with two
separate $90^{\text{o}}$ DW nucleations at $H_{c2}$ and $H_{c4}$, respectively. When $45^{\text{o}}<\phi<90^{\text{o}}$ and
$K_{u1}>\epsilon_{90^{\text{o}}}$, the magnetization switches from [0$\overset{-}{1}$0] to [$\overset{-}{1}$00] at $H_{c1}$, where $\Delta
E_{270^{\text{o}}\rightarrow180^{\text{o}}}=\epsilon_{90^{\text{o}}}$. Because $\Delta E_{90^{\text{o}}\rightarrow0^{\text{o}}}$\ decreases with
increasing applied field and becomes $\epsilon_{90^{\text{o}}}$ at $H_{c4}$ (see Fig.~3(b)), two successive DW nucleations appear for
$H_{c2}<H_{c4}$. Upon further increasing the field, $\Delta E_{90^{\text{o}}\rightarrow0^{\text{o}}}$ becomes negative, and finally reaches
$-\epsilon_{90^{\text{o}}}$ at $H_{c3}$, where the magnetization switches backwards from [100] to [010]. As a result, the magnetization loops
contain three steps. For $H_{c2}>H_{c4}$, the domains aligned along $90^{\text{o}}$ are energetically stable when the applied field exceeds
$H_{c2}$, resulting in two-step loops.

By comparing the expressions for $H_{c2}$ and $H_{c4}$ the field orientation for the occurrence of one-step or three-step loops can be obtained.
For the simple case $\delta=0$, the condition $\tan\phi<K_{u1}/\epsilon_{90^{\text{o}}}$ needs to be satisfied, where $0<\phi<45^{\text{o}}$ for
the one-step loops and $45^{\text{o}}<\phi<90^{\text{o}}$ for the three-step loops, respectively. Our model predicts that the ranges of angles
for which a one-step loop should be observed are $-45^{\text{o}}<\phi<45^{\text{o}}$, $-28^{\text{o}}<\phi<28^{\text{o}}$, and
$-43^{\text{o}}<\phi<36^{\text{o}}$ for samples A, B, and C, respectively. The critical angles separating the occurrence of two-step and
three-step loops are $\phi=90^{\text{o}}\pm13^{\text{o}}$ for sample A. Our model calculations nicely agree with experiment.

Introducing two successive DW nucleations allows us to consistently interpret the $180^{\text{o}}$ magnetization reorientation. Since the DWs
induced during the first nucleation are energetically unstable, it should be very hard to observe these intermediate domains. In real films,
however, DW nucleation and propagation are often perturbed by defects and roughness~\cite{NM-2-521}, implying magnetic switching is not as sharp
as predicted. As illustrated in Fig.~4(a), we observed the metastable intermediate states in the loops of sample A close to the critical angles
separating the occurrence of two-step and three-step loops. The blue curve (increasing field) reveals an overshoot for magnetic switching from
[$\overset{-}{1}$00] to [100], indicating that the Fe spins align for a short time along [010] before jumping to [100]. The red curve
(decreasing field) reveals a similar feature. We note that the second intermediate state in the blue curve in Fig.~4(a) is not collinear with
the first intermediate state in the red curve and vice versa, which is different from the loop shown in Fig.~1(d). The non-collinearity implies
that not all spins switch from [$\overset{-}{1}$00] to [010] and then to [100], but some of the spins remain aligned along [010]. This points to
coexistence of magnetic switching processes with both two-step and three-step loops. The overshoot can still be observed a few degrees away from
the critical angles. In sample B we observe a mixture of one-step and two-step loops, as illustrated in Fig.~4(b). The red curve of the
transverse MOKE loop reveals two separate $90^{\text{o}}$ DW nucleations, while the blue curve corresponds according to our model to two
successive and indistinguishable DW nucleations. Experimentally, the transition between two reversal mechanisms does not occur at one critical
angle as predicted by theory, but extends over a small finite range of angles. Time resolved MOKE may be able to detect the ultrafast
magnetization dynamics and the intermediate domain formation in the process of two successive DW nucleations.

\begin{acknowledgments}
Financial support was provided by the Fund for Scientific Research - Flanders
(FWO) as well as by the Flemish Concerted Action (GOA) and the Belgian
Interuniversity Attraction Poles (IAP) research programs.
\end{acknowledgments}

\section{Figure Captions}

FIG. 1. (Color online) (a) Definition of the angles that are used to describe
a film with in-plane cubic anisotropy and UMA. Typical longitudinal MOKE loops
for sample A with (b) one step at $\phi=8^{\text{o}}$, (c) two steps at
$\phi=88^{\text{o}}$ and (d) three steps at $\phi=68^{\text{o}}$. The blue
(red) curves are for applied fields varying from negative (positive)to
positive (negative) saturation. The arrows enclosed by a square represent the
orientation of the Fe spins.

FIG. 2. (Color online) The experimental switching fields (symbols) as a
function of the field orientation $\phi$, and the corresponding theoretical
curves for $H_{c1}$ (magenta), $H_{c2}$ (red), $H_{c3}$ (purple), $H_{c4}$
(blue), the \textquotedblleft virtual\textquotedblright\ $H_{c4}$ (dashed
gray), and $H_{c}$ (green) for (a) sample A, (b) sample B, and (c) sample C.

FIG. 3. (Color online) Energy differences $\Delta E_{180^{\text{o}}\rightarrow0^{\text{o}}}$ (red), $\Delta E_{270^{\text{o}}\rightarrow
180^{\text{o}}}$ (cyan), $\Delta E_{180^{\text{o}}\rightarrow90^{\text{o}}}$ (green), and $\Delta E_{90^{\text{o}}\rightarrow0^{\text{o}}}$
(blue) as a function of the applied field for (a) sample B at $\phi=10^{\text{o}}$ and (b) sample A at $\phi=65^{\text{o}}$.

FIG. 4. (Color online) (a) The longitudinal magnetization loop of sample A at
$\phi=76^{\text{o}}$ and (b) the transverse magnetization loop of sample B at
$\phi=29^{\text{o}}$ reveal overshoots near the critical angles for the field orientation.

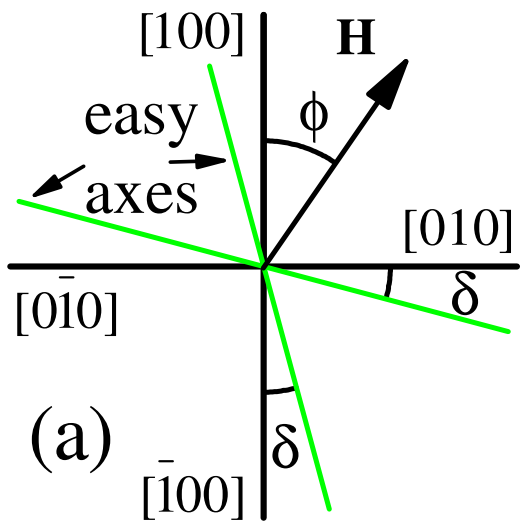
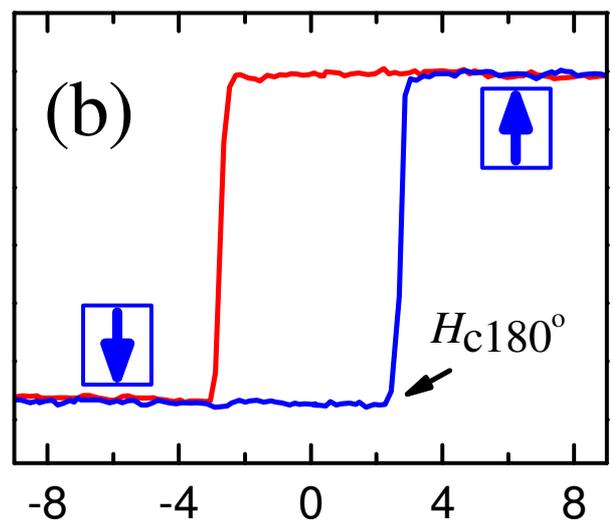
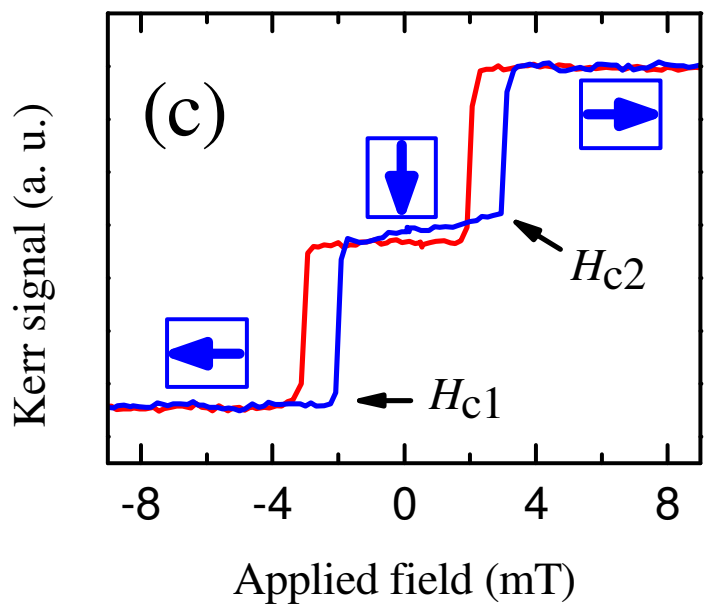
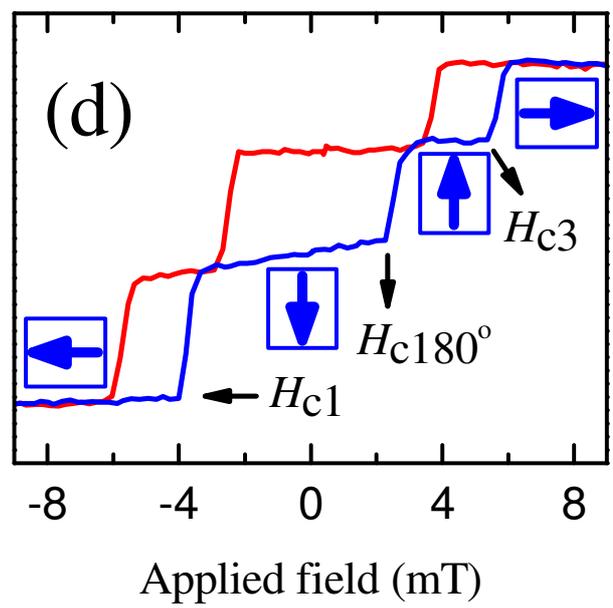
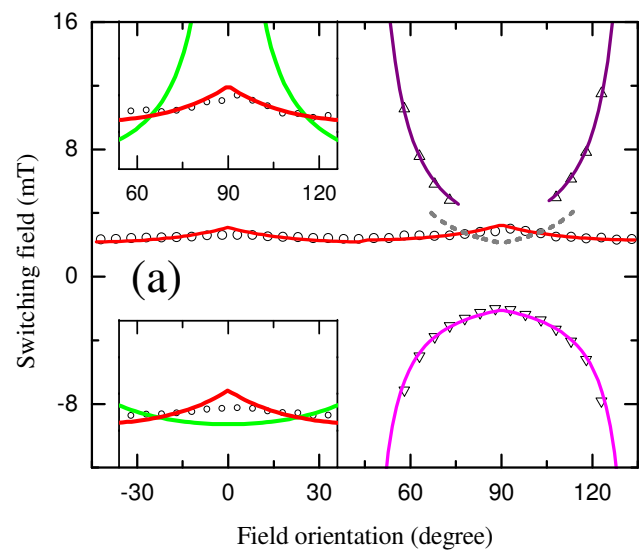
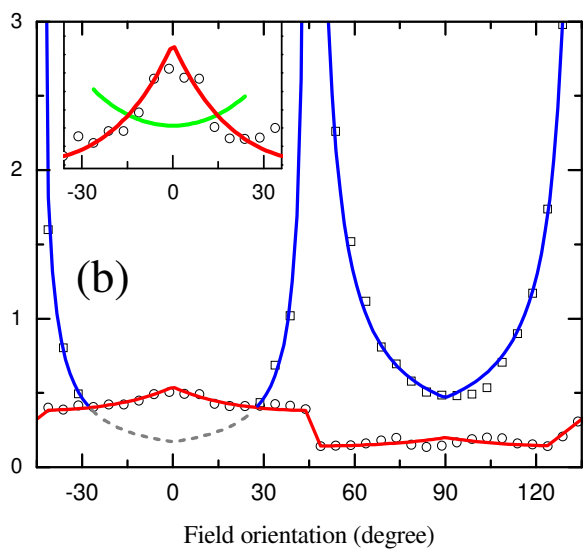
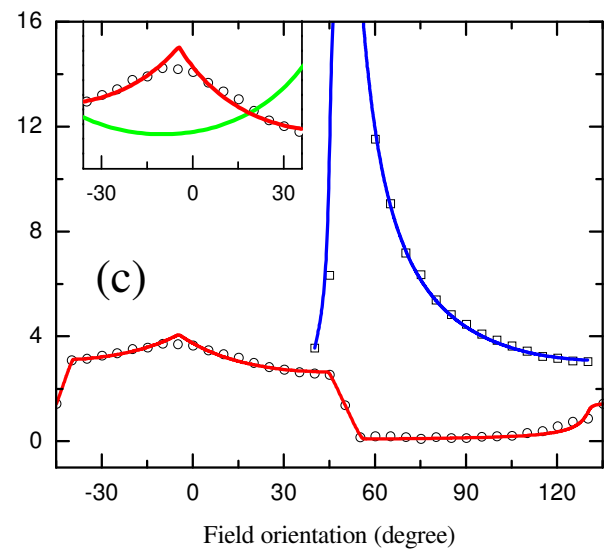
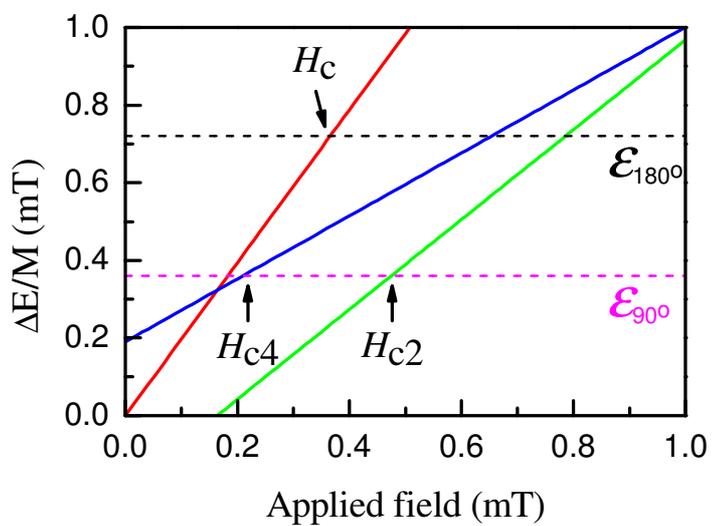
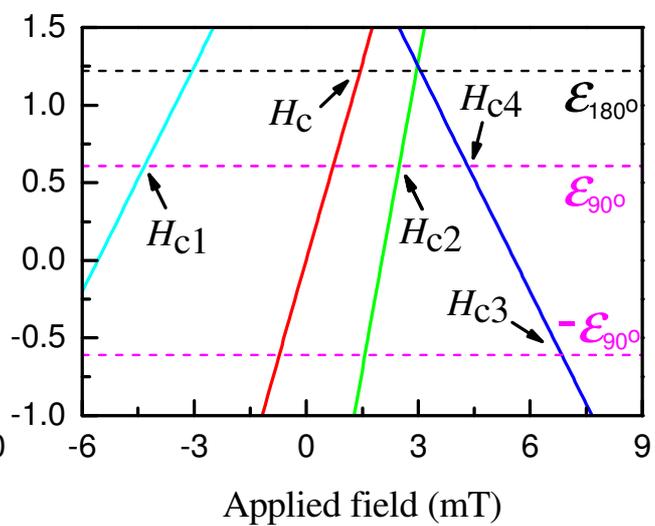
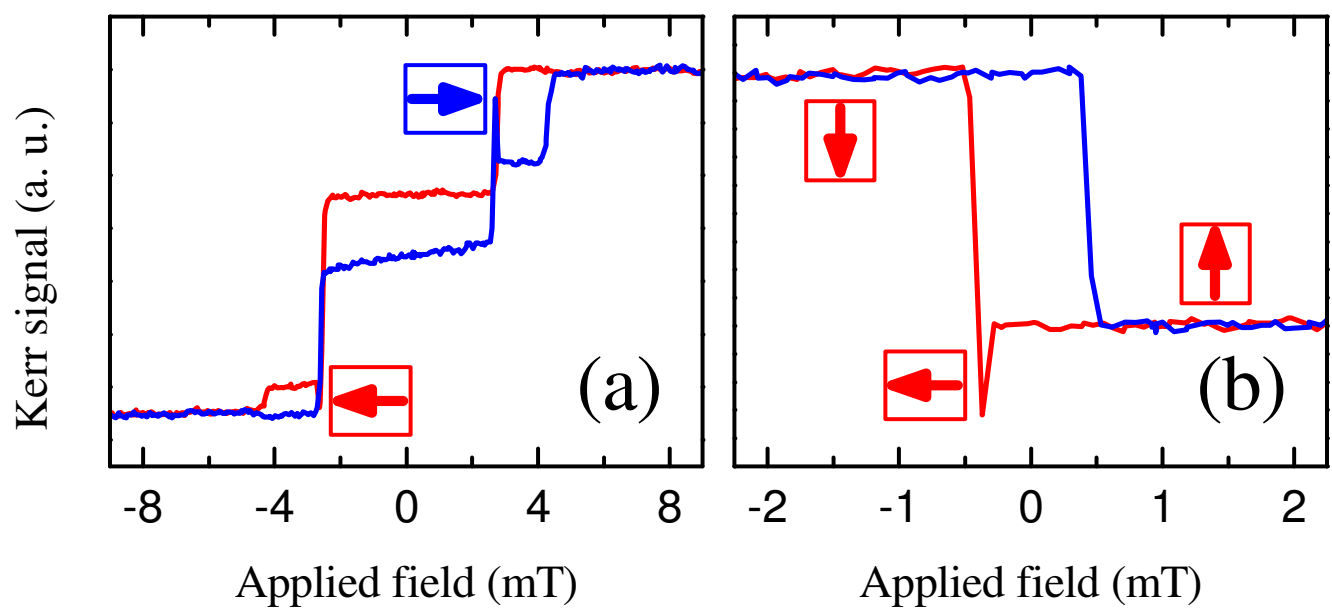
\end{document}